\def \cm{~\rm{cm}}
\def \s{~\rm{s}}
\def \km{~\rm{km}}

\def \AU{~\rm{AU}}
\def \erg{~\rm{erg}}

\def \yr{~\rm{yr}}
\def \pc{~\rm{pc}}

\def \astrobj#1{#1}
\documentclass[12pt,preprint]{aastex}
\usepackage{natbib}

\begin{document}
\title{MULTI-JETS STRUCTURE RATHER THAN ROTATION OF THE NGC 1333 IRAS 4A2 PROTOSTELLAR JETS}

\author{Noam Soker\altaffilmark{1} \& Liron Mcley\altaffilmark{1}}

\altaffiltext{1}{Dept. of Physics, Technion, Haifa 32000, Israel;
soker@physics.technion.ac.il.}

\begin{abstract}
We analyze the velocity gradient across the jets of the young stellar object (YSO) \astrobj{NGC 1333 IRAS 4A2} and explain it
by decomposing the two opposite jets to two opposite sub-jets.
The two sub-jets have the same radial velocity and the angle between them is only several degrees.
Each of the two sub-jets is composed of two components.
We also show that the alternative interpretation of jets' rotation is unlikely to account for the velocity gradient.
The two sub-jets directions are constant, and there is no indication for precession.
The line connecting the centers of the sub-jets defines a preferred direction in a plane parallel to the accretion disk launching the jets.
We suggest that the preferred constant direction in the accretion disk is determined by the semi-major axis of a highly eccentric
orbit of a brown dwarf or a massive planet companion.
At each periastron passage the companion perturbs the accretion disk with a quadrupole mode, e.g., two spiral arms,
that lead to the launching of the two sub-jets.
We predict that a careful monitoring of the system will reveal a periodic activity with a period of few months to several years.
\end{abstract}

\section{INTRODUCTION}
\label{sec:intro}

Asymmetry in the line of sight velocity across some jets launched by young stellar objects (YSOs)
has been interpreted as caused by a large scale rotation of the material in the jets
(e.g., \citealt{Bacciotti2002}; \citealt{Coffey2004, Coffey2007}; \citealt{Lee2007}, \citeyear{Lee2009};
\citealt{Chrysostomou2008}; \citealt{Choi2011a}).
These measurements, however, are quite challenging and complicated (e.g., \citealt{Coffey2012}).
In two earlier papers (\citealt{Soker2005},  \citeyear{Soker2007}) we analyzed the results of
\citet{Bacciotti2002}, \citet{Coffey2004} and \citet{Coffey2007} and argued that the observations
do not support, and even contradict, the interpretation of jets rotating around their symmetry axes.
The claims for jets' rotation in the YSOs  HH~211 (\citet{Lee2007}, \citeyear{Lee2009})
and Ori-S6 \citep{Zapata2010} were found to be quite problematic as well \citep{Soker2010}.

Fast rotation around the jet axis is predicted by the magneto-centrifugal acceleration (MCA) model
for jet launching (e.g., \citealt{Anderson2003, Ferreira2006}).
In this model the magnetic fields that are anchored into the accretion disk-star system play a dominate
role in accelerating the jet's material from the accretion disk by a lever-arm torque.
In this model the specific angular momentum carried by the jet's material is $\sim (\Lambda R_f)^2 \Omega_f$,
where $R_f$ is the radius at the \emph{foot-point} of the jet on the disk, $\Omega_f$ is the angular velocity at $R_f$,
and $\Lambda > 1$ is the lever-arm.

Instead of the rotation interpretation, \citet{Soker2005} proposed that interaction of the jets with a
twisted-tilted (wrapped) accretion disk can form the observed asymmetry in the jets' line of sight velocity profiles.
The jets interact with the ambient gas residing on the two sides of the disk,
e.g., a weak disk outflow or corona.
Such a model can explain the asymmetry in the line of sight velocity across the jets, avoiding
the need to invoke jet rotation with a huge amount of angular momentum.

Another model that can account for such observations without invoking jets' rotation was
proposed by \citet{Cerqueira2006} who assumed a precessing jet whose ejection velocity
changes periodically with a period equals to the precession period.
Practically, the dependance of the jet's expansion velocity on direction around the
symmetry axis leads to the same effect as the model of \citet{Soker2005}.

In a recent paper \citet{Choi2011a} argue that the velocity gradient across the jets of the YSO
\astrobj{NGC~1333~IRAS~4A2} can be interpreted as jets' rotation, and that this supports the
disk-wind models \citep{Pudritz2007}.
In section \ref{sec:properties} we summarize the properties of the outflow from NGC~1333~IRAS~4A2, and in
section \ref{sec:4jets} we present our two sub-jets interpretation.
Our short summary is in section \ref{sec:summary}.

\section{THE OUTFLOW FROM NGC~1333~IRAS~4A2}
\label{sec:properties}

The following properties of the jets of NGC~1333~IRAS~4A2 (hereafter IRAS~4A2) are relevant to us.
The central star has a mass of $M_\ast=0.08 M_\odot$ and it accretes mass at a rate of
$\dot M_{\rm acc} = 1.6 \pm 0.9 \times 10^{-6} M_\odot \yr^{-1}$ \citep{Choi2010}.
\citet{Choi2011a} assume that the velocity of the jet they observe in SiO is as that of
the parts observed in molecular hydrogen line by \citet{Choi2006}: $71 \km \s^{-1}$ in the plane of the sky. From
that and the assumption that the velocity gradient is due to rotation
they derive the specific angular momentum of the gas in the jets to be $l_j\simeq 1-2 \times 10^{21} \cm^2 \s^{-1}$.
Assuming the magnetohydrodynamic wind model for the rotating jets (\citealt{Anderson2003}; \citealt{Pudritz2007})
they calculate the jets' foot-point to be at $R_f \simeq 2 AU$.
At the location of this foot-point the specific angular momentum of the disk material is
\begin{equation}
l_d(R_f) = 1.8 \times 10^{19} \left( \frac{M_\ast}{0.08M_\odot} \right)^{1/2}
      \left( \frac{R_f}{2 \AU} \right)^{1/2}  \cm^{2} \s^{-1} \simeq 0.01 {\it l_j},
\label{eq:ld}
\end{equation}
and its specific kinetic energy is
\begin{equation}
e_d = 1.8 \times 10^{11} \left( \frac{M_\ast}{0.08M_\odot} \right)
      \left( \frac{R_f}{2 \AU} \right)^{-1} \erg  \simeq 3.6 \times 10^{-3} {\it e_j},
\label{eq:ed}
\end{equation}
where $e_j \simeq (70 \km \s^{-1})^2/2$ is the specific kinetic energy of the gas in the two jets.

The above relations have two implications.
First, the `lever-arm' is $\Lambda = [l_j/l_d(R_f)]^{1/2} \simeq 10$ to $\Lambda = [e_j/e_d(R_f)]^{1/2} \simeq 17$.
The lever-arm is defined as the ratio of the Alfven radius, inward to which the outflowing gas
is controlled by the disk's magnetic field, to the foot-point radius.
Although values of $\Lambda >10$ were suggested in the literature
(e..g, \citealt{Casse2000}, who by `lever-arm' refer to a different quantity) and cannot be ruled out,
this is a very large value for $\Lambda$ which has a typical value of $2 \la \Lambda \la 3$
(e.g., \citealt{Anderson2003}; \citealt{Pudritz2007}).
We also note that the outflow velocity of $v_j \sim 70 \km \s^{-1}$ is about the escape velocity from a YSO
of mass $M_\ast=0.08 M_\odot$, suggesting to us a jet origin very close to the star.

Second, angular momentum and energy conservation with the above ratios imply that the jets mass outflow rate
is limited to
$\dot M_j < 0.0036 \dot M_{\rm acc} \simeq 6 \times 10^{-9} M_\odot \yr^{-1}$.
There is no reliable derivation of the mass loss from IRAS~4A2.
Scaling the results of \citet{Blake1995} to a distance of $235 \pc$ and an age of $\sim 50,000 \yr$ \citep{Choi2011a},
gives a mass outflow rate of $\sim 10^{-6} M_\odot \yr^{-1}$.
\citet{Lefloch1998} finds the number density in the wings of high-velocity components
of the outflow, $v \simeq 30 \km \s^{-1}$, to be $\sim 0.6-2\times 10^6 \cm^{-3}$ close to the star.
For jets' radius of $1^{\prime \prime}$ the calculated mass loss rate is $\sim 10^{-5} M_\odot \yr^{-1}$.
This seems to be too high.
The conclusion here is that although better determination of the mass outflow rate is required,
available data, with their large uncertainties, do suggest that the mass outflow rate is larger than the
mass outflow rate allowed in the rotating jet model.

\begin{figure}[b]
\begin{center}
\includegraphics[scale=0.7]{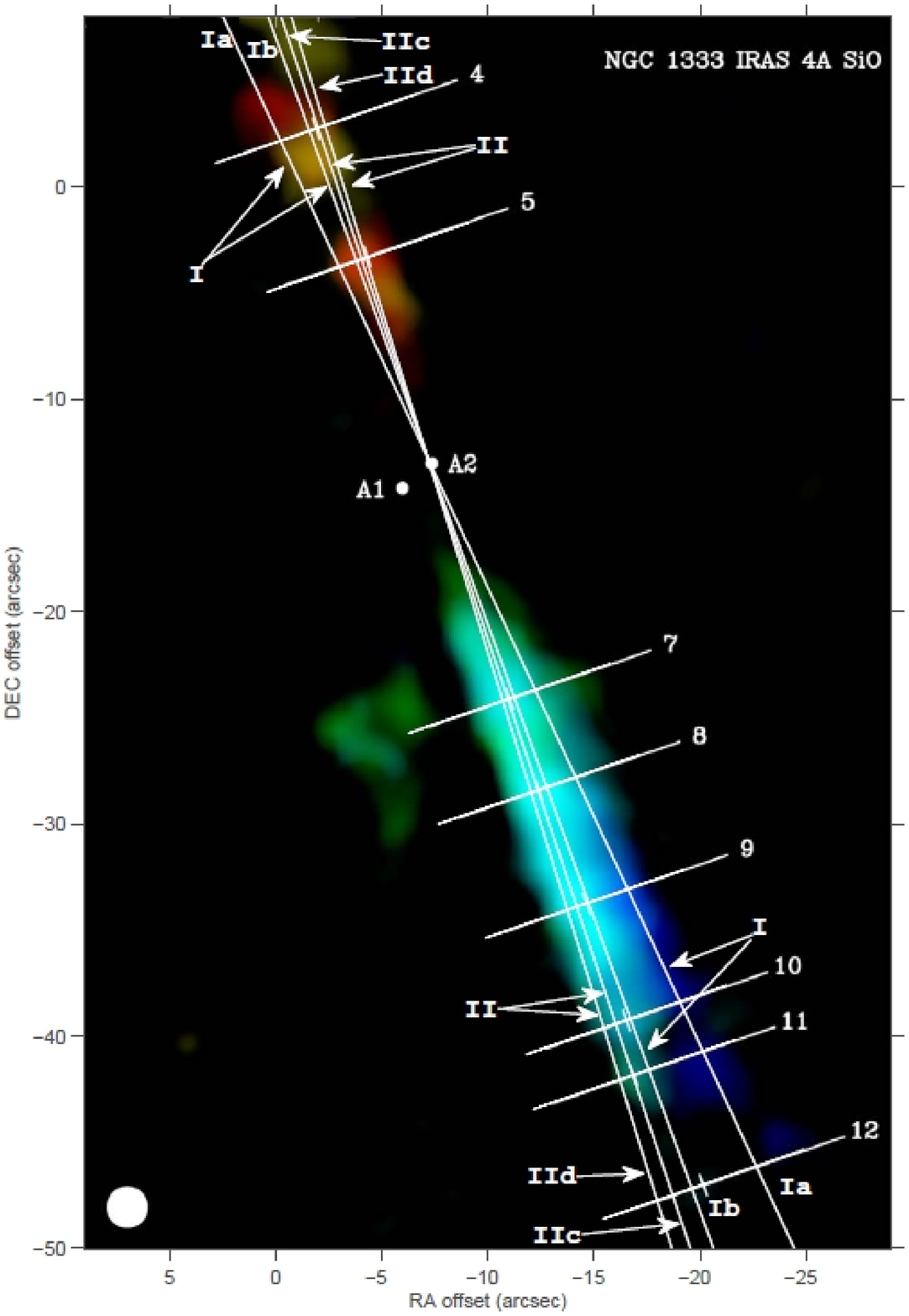}
        \caption{The color composite image of the NGC 1333 IRAS 4A2 bipolar jet in the SiO
$v=0\;J=1\rightarrow0$ line from \citet{Choi2011a}.
The white lines across the jets that are labelled by numbers, the `cuts', are from the original figure of \citet{Choi2011a}, while the
four lines along the two opposite jets are our addition.
These four lines mark the two opposite sub-jets ($I$ and $II$), each composed of two components that we identify in the sub-jets (see text).       }
   \label{fig:jets}
\end{center}
\end{figure}

\cite{Soker2005} proposed that interaction of the jets with a twisted-tilted (wrapped) accretion disk can form the observed
asymmetry in the jets line of sight velocity profiles.
This type of interaction predicts that the western edge of the northern jet be symmetric to the eastern edge of the southern jet.
Indeed, as seen in Figure \ref{fig:jets} the eastern edge of the northern jet (left side of upper jet in the figure) and the
western edge of the southern jet (right side of the lower jet in the figure) have wiggling boundaries, while
the western edge of the northern jet and the eastern edge of the southern jet have relatively sharp boundaries \citep{Choi2005, Choi2011a}.
This not only shows the type of symmetry expected from the model of \cite{Soker2005}, but also that an interaction with the
surrounding gas might account for the velocity gradient
perpendicular to the jets' axis.
In the next section we examine in greater detail the jets, and propose yet another possible process that operates in the
launching process of the IRAS~4A2 jets.


\section{A DOUBLE-JET MODEL FOR THE JETS OF NGC~1333~IRAS~4A2}
\label{sec:4jets}

\begin{figure}[b]
\begin{center}
\includegraphics[scale=0.7]{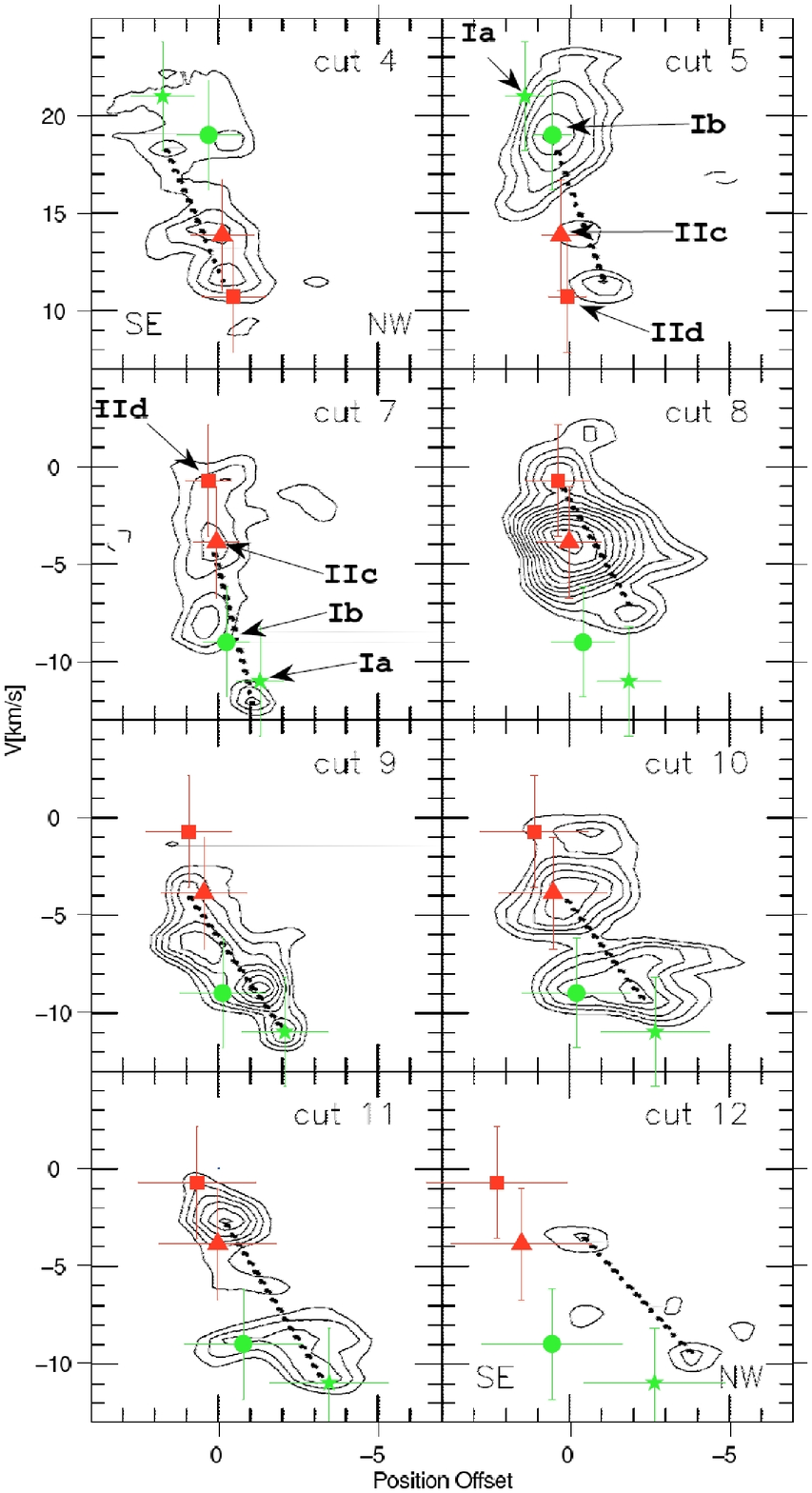}
        \caption{The position-velocity (PV) diagrams from \citet{Choi2011a}.
        The colored crosses are our addition.
        The center of each cross represents the Doppler shift and position of a sub-jet component along the appropriate cut.
        Each of the four symbols of the crosses represents one component, as marked on two of the panels.
        }
   \label{fig:PV1}
\end{center}
\end{figure}

\citet{Choi2011a} used the Very Large Array to observe the bipolar jets of the protostar IRAS~4A2
in the SiO $v = 0 \; J = 1\rightarrow 0$ line.
They investigated the kinematics of the two opposite SiO jets by placing
slits perpendicular to the jets' axis; these were termed cuts.
The jets and the positions of the cuts (marked by numbers) are shown in Figure \ref{fig:jets}.
\citet{Choi2011a} produced position-velocity (PV) diagrams for each cut, where
they positioned the zero displacement of each cut at the emission peak according to \cite{Choi2005}.
The PV diagram for each of the perpendicular slits is presented in Figure \ref{fig:PV1} here, taken from Figure 2 of \citet{Choi2011a}, where
the colored crosses are our addition which we discuss below.


Noticing that the sense of velocity gradient is the same in all cuts they argued for jets' rotation.
To demonstrate their interpretation that the velocity gradient is due to jets' rotation, they drew
a velocity gradient for each cut; these are the black dotted lines on the PV diagrams of Figures \ref{fig:PV1}.
We criticized this interpretation in section \ref{sec:properties}.
We hence aim at a different interpretation of the velocity structure.
Examining the PV diagram of the different cuts, we notice that there is a multi-peak structure in every one of them.
(1) There are two clear separated structures in cuts 10 and 11.
(2) There are three or four peaks in most cuts.
(3) In cut 8 there are only three peaks, but the central one is the strongest among all peaks.
This is as if two peaks overlap each other.

We try to fit the peaks with the following structure.
We assume the existence of two sub-jets, each composted of two components.
Here a sub-jet refers to both northern and southern parts. Namely, a sub-jet has two opposite parts relative to the central star.
The same holds for a 'component'.
We model each of the four components as a straight line of gas on the two opposite sides of the star,
as if it is a jet of its own.
These four components are drawn on Figure \ref{fig:jets}.
The axis of each component is parametrized by two angles.
First is the inclination relative to the plane of the sky, $\theta$.
For $\theta > 0$ the northern part points away from us, i.e., red-shifted.
Second is the angle on the plane of the sky relative to component~Ia.
The values of $\theta$ and $\phi$ are given in Table \ref{tab:Table1}.
Based on \citet{Choi2006} we take all components to have the same radial expansion velocity of $v_j = 71 \km \s^{-1}$.
We compute the PV diagram of the four components and draw them as colored crosses on Figure \ref{fig:PV1}, where the velocity and position
is at the center of the cross.
The crosses themselves represent our crude estimations of the errors in the derived values of $\theta$ and $\phi$.
From our trials of different fittings we crudely estimate the errors to be of few degrees.
\begin{table}
\centering
  \caption{The angles of each component}
    \begin{tabular}{rrrr}
          &              &  \\
    \hline
\multicolumn{1}{c}{Sub-jet}& \multicolumn{1}{c}{Component}& \multicolumn{1}{c}{$\theta$}  & \multicolumn{1}{c}{$\phi$}    \\
 \hline
 \multicolumn{1}{c}{I} & \multicolumn{1}{c}{Ia}& \multicolumn{1}{c}{$12.8^\circ$}      & \multicolumn{1}{c}{$0^\circ$}   \\
 \multicolumn{1}{c}{I} & \multicolumn{1}{c}{Ib}& \multicolumn{1}{c}{$11.2^\circ$}      & \multicolumn{1}{c}{$5.0^\circ$} \\
 \multicolumn{1}{c}{II}& \multicolumn{1}{c}{IIc}& \multicolumn{1}{c}{$4.6^\circ$}      & \multicolumn{1}{c}{$6.5^\circ$} \\
 \multicolumn{1}{c}{II}& \multicolumn{1}{c}{IId}& \multicolumn{1}{c}{$7.1^\circ$}      & \multicolumn{1}{c}{$7.7^\circ$} \\
 \hline
   \end{tabular}
  \label{tab:Table1}
\end{table}

The four components do not perfectly fit the observed PV diagram as can be seen in Figure \ref{fig:PV1}.
First, there is a substantial interaction of the jets with the ambient gas. This is clearly evident from the sharp bend of the northern jet
which occurs further out (\citealt{Choi2006}; not seen here).
Other regions also show some interaction with the surroundings, as evident from the wiggling boundaries of the jets
(see discussion in section \ref{sec:properties} here).
Second, this interaction or jittering of the jets' source might cause random change in the expansion directions of the different components.

\begin{figure}[b]
\begin{center}
\includegraphics[scale=0.7]{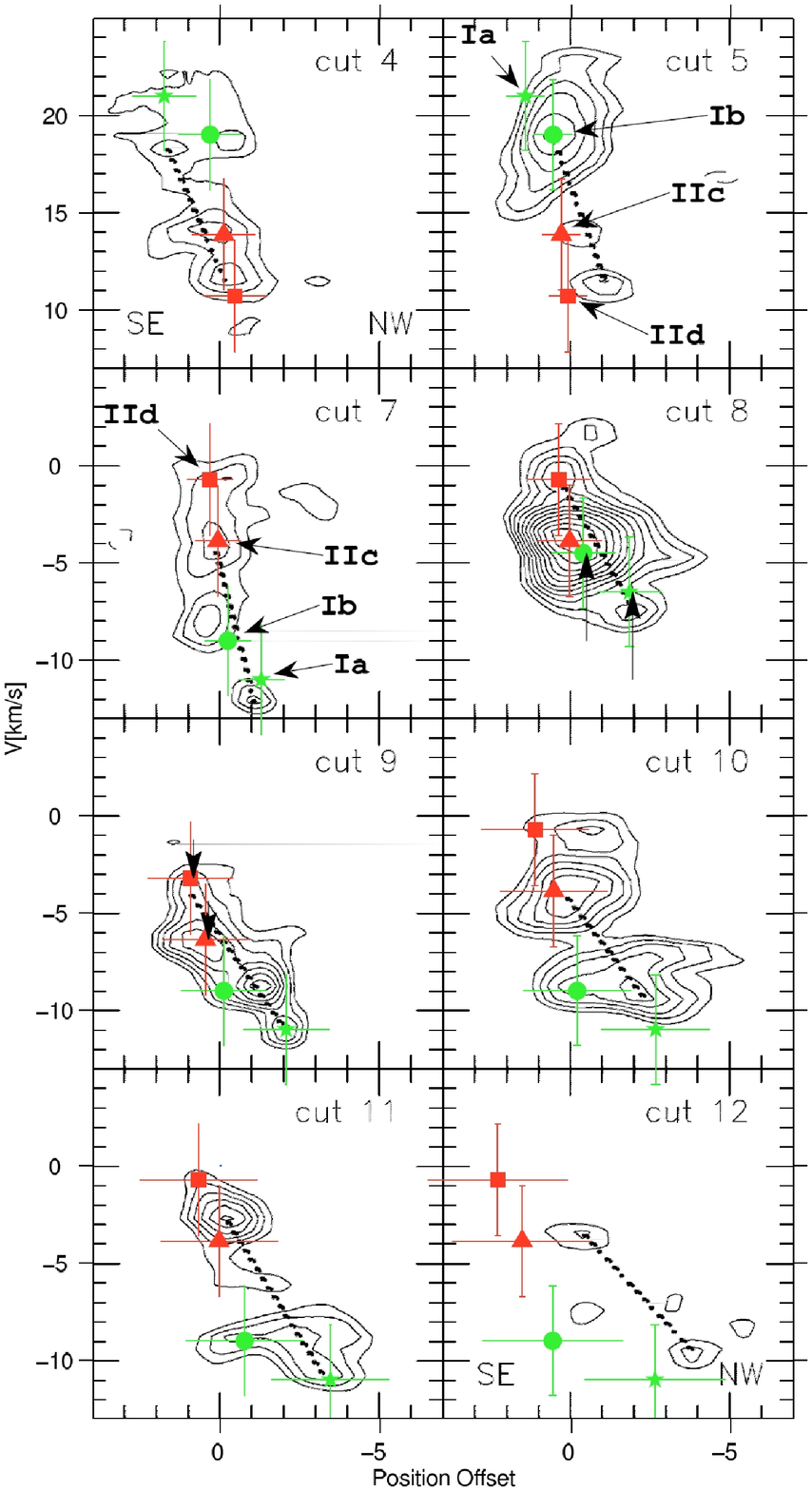}
        \caption{Like Figure \ref{fig:PV1}, but with a displacement of sub-jet I in cut 8 and of sub-jet II in cut 9, as marked
        by the vertical black arrows (see text).       }
   \label{fig:PV2}
\end{center}
\end{figure}

We notice that we can substantially improve the fitting by displacing components Ia and Ib together in cut 8,
and components IIc and IId in cut 9.
The displacement is $\Delta \theta = -3.7^\circ$, namely redward in the southern jet, for components Ia and Ib in cut 8,
and $\Delta \theta = 2.0^\circ$, namely blueward in the southern jet,  for components Ic and Id in cut 9.
The PV diagram with these displacements is presented in Figure \ref{fig:PV2}.
The displacement in cut 8 explains why the peak there is as twice as high as in the other cuts.
As the simultaneous displacement of components Ia and Ib accounts quite well for the PV diagram in cut 8, we identify
them as a single sub-jet I.
As well, we identify components IIc+IId as sub-jet II.

Our main conclusions from the simple fitting, although not perfect, are as follows.
\begin{enumerate}
\item The two opposite jets of IRAS~4A2 are actually composed of two bipolar sub-jets.
Each of the bipolar sub-jets is composed of two components.
\item We can make the fitting with two sub-jets that have the same radial velocity, and with a very small angle between them.
These suggest a common launching site.
\item There is no sign of precession. The sub-jets and the two components composing each sub-jet maintain an average constant direction.
\item The structure of four-components and two sub-jets exists on both sides of the star (north and south jets).
This shows that the sub-jets structure does not result from a random collision and stochastic interaction with the surrounding gas.
\item The relatively large displacement of one sub-jet we identify in cuts 8 and 9, according to our interpretation,
can result from either change in launching direction or from interaction with the surrounding gas.
The interaction with the surrounding gas can explain the small displacements seen in the different cuts and in small changes in structure
and intensities.
However, the large displacement found in cuts 8 and 9 are more likely to result from instabilities and stochastic behavior of the launching process
that lead to a temporary change in the velocity (or launching angle) of one sub-jet.
As well, the existence of the two sub-jets also points to a complicated launching process. We speculate on a possible cause in the next section.
\item No extra component of jet rotation is required in the two sub-jets interpretation.
\end{enumerate}

We note that the sub-jets structure is found in the PV diagram and not in the image.
This is possible because the sub-jets are almost in the plane of the sky, i.e., $\sin \theta \ll 1$.
For higher values of $\sin \theta \ge 0.5$ the contribution of the radial velocity to the Doppler shift is higher, and changes
in the radial velocities of the sub-jets as they expand might overwhelm the differences between the sub-jets and components.

\section{DISCUSSION AND SUMMARY}
\label{sec:summary}

Our interpretation of two sub-jets structure as presented in the previous section requires that the launching process has a preferred direction that
does not change with time, at least not for the time period span by the jets of $\sim 500 \yr$.
This excludes the possibility that the preferred direction is the direction toward a companion in a circular orbit.
This is because for a slowly varying direction toward the companion the orbital period must be thousands of years or more.
Such a companion will be at a too large distance to influence the launching process of the jets.
A close companion changes its direction on a short time scale.

However, if the orbit is eccentric the semi-major axis of the orbit defines a preferred direction that can maintain
a constant direction for a very long time.
We propose, therefore, that the launching process of the jets in IRAS~4A2 is substantially influenced by the presence
of a companion on a highly eccentric orbit.

\cite{Ardila2005} present the interaction of a disk with a close binary star on a parabolic orbit.
This passage creates two large-scale spiral arms in the disk.
The spiral arms extend inward to a distance of $\sim 0.1 r_p$, where $r_p$ is the periastron distance,
and preserve a more or less constant direction for most of the time.
The dense-spiral arms structure in the gas component of the disk lasts for a time of $\sim 0.5 P_p$, where $P_p$ is the orbital period of a
circular orbit of radius $r_p$.
The two spiral arms can lead to the formation of the two sub-jets we identify in the outflow from IRAS~4A2.

We can estimate plausible parameters for the orbit.
Taking the IRAS~4A2 jets' velocity of $70 \km \s^{-1}$ to be the escape speed from their launching radius $r_L$,
we find $r_L \simeq 6 R_\odot$, where the mass of the star is $M_\ast=0.08 M_\odot$ \citep{Choi2011b}.
Let the companion periastron be then at $r_p \simeq 3-10 r_L \simeq 20-60 R_\odot$, and the apastron be at $\xi \gg 1$ times this distance.
The orbital time of the binary system (assuming the companion to have a mass of $M_2 << M_1=0.08M_\odot$) is
\begin{equation}
P_{\rm orb} = 1.1
\left( \frac{M_\ast}{0.08M_\odot} \right)^{-1/2}
\left( \frac{r_p}{50 R_\odot} \right)^{3/2}
\left( \frac{1+\xi}{4} \right)^{3/2} \yr.
\label{eq:Porb}
\end{equation}
Each jets' launching episode lasts for a few months.
Over the hundreds of years life time of the jets, the many launching episodes will be smeared to continuous jets.
While the system is near apastron, the disk near the primary star can rebuild itself.

One prediction of the model we proposed is that a careful monitoring of the system will reveal a periodic activity with a period time
between few months and several years.
Observations should be ``lucky'' enough to catch the system near a periastron passage, when massive launching takes place.
The companion is a brown dwarf or a massive planet.

We cannot elaborate on how the two spiral arms in the disk lead to the formation of two sub-jets.
Simply, there is no agreed-upon model for the launching of jets that we can use to derive the process.
As for the two components of each sub-jet, numerical simulations of the gravitational interaction of the companion with the
disk around the primary star should be carried out, probably in three-dimensions. Such simulations can reveal the presence of
higher harmonics in the accretion disk in addition to the spiral arms, as well as the development of instabilities. But again,
the way the structure in the disk influences the jets formation is a more complicated process.

The main claim of this paper is that the velocity gradient across the jets of the YSO IRAS~4A2 cannot be accounted for by jets' rotation
(section \ref{sec:properties}).
They are much better interpreted as arising from a sub-jets structure (section \ref{sec:4jets}).
A similar interpretation can hold for the outflow HH 797 located in the IC 348 cluster in Perseus.
In a recent paper \cite{Pech2012} interpret the velocity gradient across the jet of this outflow
as evidence for jet rotation. We suggest that a careful examination might reveal a sub-jets structure in that outflow.
However, in the case of HH 797 interaction with the ambient gas seems to play a major role, and interpretation is more complicated.
In any case, a two sub-jets structure is clearly seen in their PV diagrams.
The origin of the two sub-jets structure might be the presence of a close companion on a highly eccentric orbit.

This research was supported by the Asher Fund for Space Research and the E. and J. Bishop Research Fund at the Technion.

\end{document}